# Issuer-Sovereign Agentic Payments

*A method for enforcing agent spending limits inside the issuer's authentication step*

Rajneesh Kaushal
rajneesh.kaushal@thalesgroup.com

Dishant Sharma
dishant.sharma@thalesgroup.com

Ashu Kanaujia
ashu.kanaujia@thalesgroup.com

## Abstract

*AI agents are beginning to make real payments. Current approaches let an agent pay by relying on a credential provider that, in the approaches deployed today, typically sits outside the cardholder's bank. The spending rules are then enforced by the card network or that provider, and not by the bank itself. This leaves the issuing bank, which carries the financial risk, with little direct control at the moment a payment happens. This paper describes Issuer-Sovereign Agentic Payments, a method that keeps that control with the issuer. The cardholder approves a spending rule once, and the bank's own authentication component records it. Later, when the agent pays a specific merchant, the bank checks the merchant .the approved rule and generates the card authentication value only if the merchant is allowed. The payment then travels the normal card rails and is validated by the issuer, with no extra dependency introduced at execution.*

## 1. Introduction

Software agents are moving from giving advice to taking actions, and one of those actions is paying for things. To support this, the payments industry has introduced protocols and products that let an agent complete a purchase using authority the user granted in advance. Well known examples include Google's AP2 [1], Mastercard's Agent Pay [2], and Visa's Intelligent Commerce [3]. These generally support two modes. In the first the user is present and approves the payment as it happens. In the second the user grants permission ahead of time and is not present when the payment is made. The method here applies to both, and uses the second as its main example, because it is the harder one to secure and the one where the gap in today's approaches is clearest. It enforces the user's spending limits inside the issuing bank's own authentication step, rather than depending on a party outside the bank at the moment of payment.

## 2. Problem Statement

When the user is present, security is relatively simple. The person is in the loop and can approve or reject the payment as it happens. When the user is absent, that safety net is gone, and something else has to make sure the payment stays within what the user allowed.

Today's agentic payment approaches solve this with a credential provider that, in the approaches deployed today, typically sits outside the cardholder's bank. The user connects a card and grants the agent a scope, such as a spending cap, allowed merchant types, and an expiry. The card is then represented by a token, and the spending rules are attached to that token or to a signed record of the user's intent.

The important detail is what happens at payment time. In the common model the agent does not hold a ready to use credential. It holds a reference, and when the merchant is ready to charge the card, the system calls back to the credential provider to exchange that reference for the real cryptogram. Mastercard's tokenization service works this way. Tokenizing a card returns a token with a unique reference, and the requestor later uses that reference to request a token and a fresh cryptogram at the time of the transaction [6]. The cryptogram is produced by the network's token service, and the spending scope is checked at the network layer rather than at the bank.

This has two consequences. First, there is a hard dependency at the worst moment, since the payment cannot be completed unless the external provider is reached and responds at the point of sale. Second, and deeper, the issuing bank carries the financial risk if a payment goes wrong. Yet enforcement of the cardholder's spending rules sits with the network or an outside provider. The bank is tightly coupled to a third party for control over its own cardholder's spending and has little direct say at the moment of authorization.

Google's AP2, Mastercard's Agent Pay, and Visa's Intelligent Commerce each take a considered approach, and each is effective within its own design. What they share is that enforcement sits away from the issuer, and the issuer takes part mainly at final authorization, after the credential has been formed elsewhere. Even where such a provider could be hosted by the bank, the enforcement in these models is still applied around a credential that already exists, rather than deciding whether it is formed at all. This paper takes the opposite position, moving enforcement into the issuer's own authentication step, so that the bank decides whether a payment credential is created at all.

## 3. Proposed Solution

The method keeps control with the issuing bank by using a component that lives inside the bank's own authentication system. To follow how it works, it helps to be clear about two things that are already in place before the method begins.

The first is the card token. Before any agent payment, the card is tokenized through the normal route, using Visa Token Service or Mastercard MDES [5], when the user first connects the card to the agent. This is the same tokenization a merchant or wallet uses to store a card on file. It produces a network token, often called a DPAN, that stands in for the real card number, the PAN, and the agent holds it the way any card on file party would. This step is ordinary and not part of what this paper proposes, and it answers a common question, namely how the agent comes to hold the card credential in the first place.

The second is the authentication value. During card authentication under 3-D Secure [4], the issuer's Access Control Server produces a cryptographic value, the CAVV on Visa and the AAV on Mastercard, that later proves to the authorization system that the cardholder was authenticated. This value normally covers one specific merchant. The Access Control Server sits in the issuer's domain and generates this value using the issuer's key, and even when a certified vendor runs it, it does so on the bank's behalf and under the bank's keys. This is why the issuer is the natural place to enforce a spending rule, since it already owns the component that decides whether an authentication value exists.

Building on these two facts, the method adds a component inside the issuer's authentication system. For clarity we refer to it as the issuer's constraint component. The idea is straightforward. When the user gives consent, the component records the spending rule but creates no payment credential. Later, when the agent pays a specific merchant, the component checks that merchant against the recorded rule and generates the authentication value only if the merchant is allowed. The two steps, checking and generating, happen together, and they happen at the issuer.

A concrete example helps. Lena, a cardholder in France, uses an AI travel agent. On Monday she tells it to book any airline for up to 500 euros within seven days, and she approves this. On Thursday the agent finds a flight and pays, with Lena not involved. The payment to the airline succeeds because it fits her rule, while an attempt to pay an unrelated merchant, such as an electronics retailer, would not produce a usable credential at all.

The full flow is shown in Figure 1 and described below in two phases.

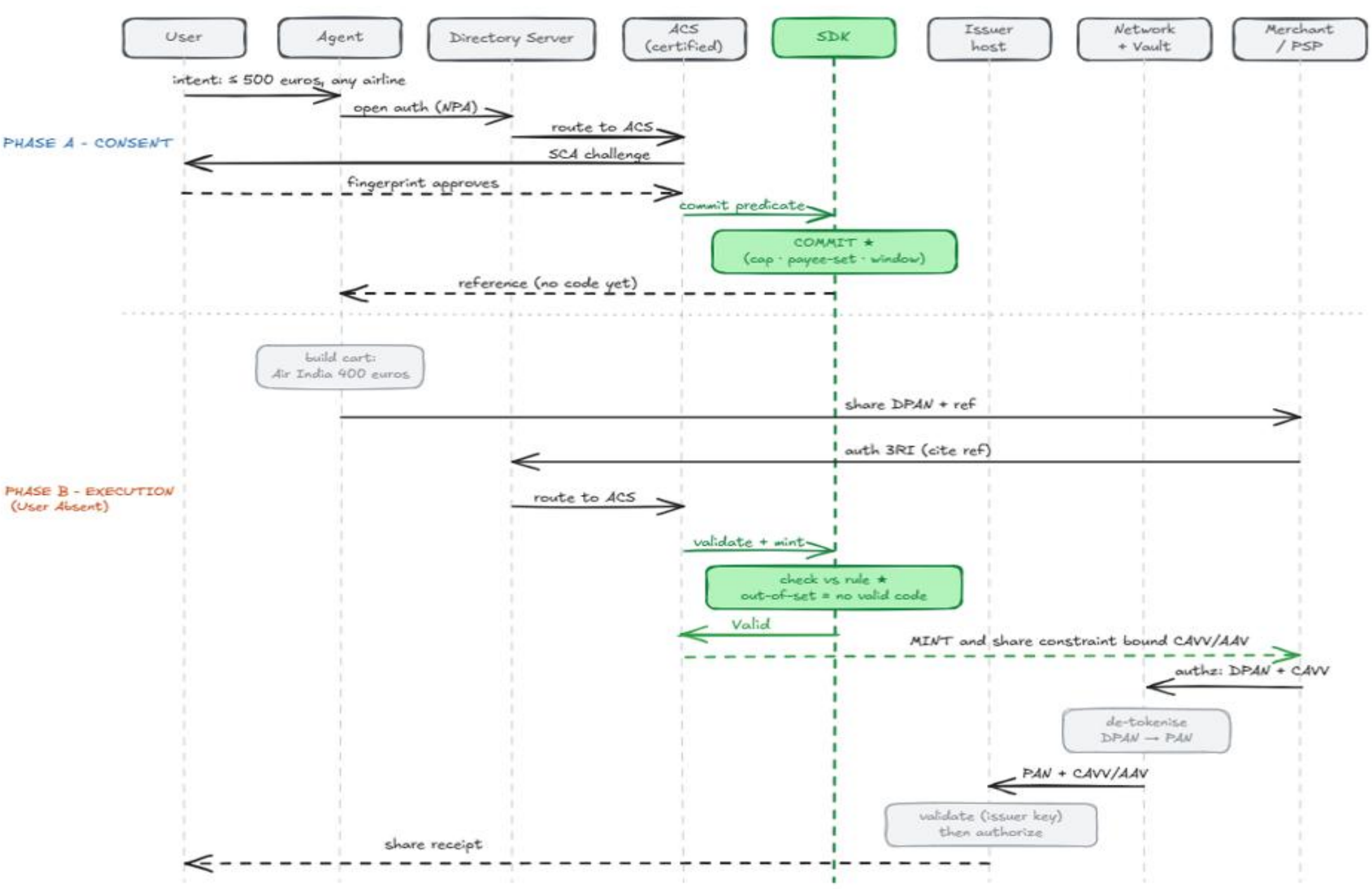

*Figure 1. End to end flow of Issuer-Sovereign Agentic Payments, showing the consent phase and the execution phase.*

Phase A is consent, and the user is present. It begins when Lena tells her agent what she wants, which is any airline for up to 500 euros within seven days. The agent opens an authentication request to the Directory Server. This request is non-payment authentication, so no charge is involved at this stage. The Directory Server routes the request to the issuer's Access Control Server. The Access Control Server then sends a challenge back to Lena to confirm that she is really the cardholder, and she approves with her fingerprint or any other suitable method. Once she is verified, the Access Control Server passes the request to the issuer's constraint component and asks it to commit the rule. The component records the rule, which holds the spending cap, the set of allowed merchants, and the time window. Notice what does not happen here. No authentication value and no payment credential are created. The component simply returns a reference that confirms the rule exists, and that reference travels back to the agent. At the end of the consent phase, all that exists is a recorded rule and a reference to it, and nothing that could be used to pay.

The recorded rule can be given stronger protection where needed. The component can commit to it as a cryptographically signed object, so that any later change to the cap, the allowed set, or the window is detectable and the approved rule can be proven in a dispute. This protects the integrity and origin of the rule and is separate from the check that decides whether an authentication value is generated.

Phase B is execution, and the user is absent. Days later the agent has found a flight and needs to pay. It begins when the shopping agent hands the network token, the DPAN, to the merchant's agent, together with the reference from the consent phase. The merchant's agent then opens an authentication request of its own, this time a request suitable for a cardholder who is not present, and it cites the reference, so the bank knows which rule applies. The request reaches the Directory

Server, which routes it to the Access Control Server as before. The Access Control Server passes it to the issuer's constraint component and asks it to validate and generate. This is the central step of the method. The component first checks the specific merchant against the recorded rule, confirming that the merchant is in the allowed set, that the amount is within the cap, and that the request falls inside the time window. Only if these checks pass, does it generate the authentication value, the CAVV or the AAV, for this transaction, and as in any card payment that value covers a single merchant. If the merchant is outside the approved set, the component will not generate a valid authentication value, and so within the system no usable payment credential exists for that transaction. The check and the generation are a single step, which means there is no window in which a credential exists for a merchant that was never approved.

The authentication value then travels back to the merchant's agent, and from this point the payment behaves like an ordinary authenticated card transaction. The merchant's agent submits the authorization to the network, carrying the network token and the authentication value. The network de-tokenizes the token, turning the DPAN back into the real card number, the PAN, and forwards the PAN and the authentication value to the issuer. The issuer validates the authentication value using its own key and authorizes the payment. There is no further call back to the constraint component at this stage, because the issuer validates the value in the usual way, exactly as it would for any authenticated card payment. Finally, the approval travels back as a receipt, and the payment is complete, having stayed within the limits Lena approved.

Two cryptographic values here are easy to confuse. The network token cryptogram belongs to the tokenization service and is checked when the token is de-tokenized, while the authentication value, the CAVV or AAV, belongs to the issuer, and this method concerns only the second.

The rule need not be limited to a single purchase. It can permit more than one payment within the cap and the window, and each payment is checked against the current state of the rule, so the cap and the window hold across everything done under one consent. The cap and the window are only two examples of what the rule can express. The same mechanism can carry other constraints checkable when the value is generated, such as allowed merchants or merchant categories, a per payment amount, a limit on the number of payments, or the permitted currency etc.

Placing the component at the Access Control Server lets it inherit an environment the issuer already trusts, since the server is certified, runs under the issuer's key management, and is already where cardholder authentication happens, so it adds no new trust boundary or holder of sensitive keys.

Two properties deserve emphasis. The spending rule is enforced at the moment the value is generated, inside the issuer, and the guarantee is stated honestly, namely that the issuer's component will not create a valid authentication value for a merchant outside the approved set, so within the system no payable credential is formed for such a merchant.

## 4. Advantages

**Zero-integration processing.** The value is generated only after the issuer checks the limits, then rides the existing rails in the standard authentication field, so every party handles it as a normal authenticated card payment with no change to processing.

**A decoupled ecosystem.** The shopping agent and the merchant's agent need no shared credential provider between them, because the bank enforces the limits directly, and they interact only to pass the token and the reference.

**Control stays with the issuer.** The party that carries the risk is the one that decides whether a payment credential is created, which aligns responsibility with authority and is the alignment missing in today's model.

**Enforcement at the point of credential generation.** The rule is checked before the credential exists and the value is generated only when the rule is satisfied, so for a merchant outside the approved set no usable credential is produced in the first place.

**Transactions look standard on the wire.** The value rides the normal field, so the transaction is indistinguishable from any other authenticated card payment on the network and does not expose the agent arrangement.

## 5. Conclusion

Agent payments are arriving quickly, and the harder case is when the user is not present to approve the purchase. Today's approaches place the spending credential and its enforcement with a party outside the cardholder's bank, which leaves the bank dependent on that party at the moment of payment and separates the control of spending from the risk of loss. This paper described a method that returns that control to the issuer. The cardholder approves a rule once, the issuer's own component records it, and the issuer generates the authentication value only when a later payment fits the rule. The payment then travels the standard card rails and is validated by the issuer in the usual way, with no extra dependency at execution, keeping enforcement where the risk already sits while leaving the rest of the ecosystem unchanged.

## References

[1] Google Cloud. Announcing Agent Payments Protocol (AP2). September 16, 2025. https://cloud.google.com/blog/products/ai-machine-learning/announcing-agents-to-payments-ap2-protocol

[2] Mastercard. Mastercard unveils Agent Pay, pioneering agentic payments technology to power commerce in the age of AI. April 29, 2025. https://www.mastercard.com/global/en/news-and-trends/press/2025/april/mastercard-unveils-agent-pay-pioneering-agentic-payments-technology-to-power-commerce-in-the-age-of-ai.html

[3] Visa. Find and Buy with AI, Visa Unveils New Era of Commerce. Visa Intelligent Commerce. April 30, 2025.

https://usa.visa.com/about-visa/newsroom/press-releases.releaseId.21361.html

[4] EMVCo. EMV 3-D Secure Protocol and Core Functions Specification.

https://www.emvco.com

[5] EMVCo. EMV Payment Tokenisation Specification, Technical Framework. https://www.emvco.com

[6] Mastercard Developers. MDES for Merchants, Digital Enablement API documentation. https://developer.mastercard.com/mdes-digital-enablement/documentation/use-cases/mdes-for-merchants-use-cases/